\documentclass[aps,prl,twocolumn,superscriptaddress,showpacs]{revtex4}

\usepackage{graphicx}

\begin{document}

\title{Computational complexity arising from degree correlations
	in networks }

\author{Alexei V\'azquez}
\affiliation{INFN and International School for Advanced Studies, Via 
Beirut 4, 34014 Trieste, Italy}
\author{Martin Weigt}
\affiliation{Institute for Theoretical Physics, University of
G\"ottingen, Bunsenstr. 9, 37073 G\"ottingen, Germany}
\affiliation{The Abdus Salam International Center for Theoretical 
Physics, P.O. Box 586, 34100 Trieste, Italy}

\date{\today}

\begin{abstract}
We apply a Bethe-Peierls approach to statistical-mechanics models
defined on random networks of arbitrary degree distribution
and arbitrary correlations between the degrees of neighboring
vertices. Using the NP-hard optimization problem of finding minimal
vertex covers on these graphs, we show that such correlations may lead
to a qualitatively different solution structure as compared to
uncorrelated networks. This results in a higher complexity of the
network in a computational sense: Simple heuristic algorithms fail to
find a minimal vertex cover in the highly correlated case, whereas
uncorrelated networks seem to be simple from the point of view of
combinatorial optimization.
\end{abstract}


\maketitle

The last few years have seen a great advance in the study of complex
networks~\cite{albert02}, where the term complex refers to the
existence of one or more of the following properties: small world
effect~\cite{watts98}, power-law degree distribution~\cite{albert02},
and more recently also correlations
\cite{redner,pastor01,berg,newman02}. On the other hand, if we focus on the
solution of a given task on top of these networks, the term complex is
better associated with the time required to solve it, i.e. with its
computational complexity~\cite{garey79}. In this context, a problem is
complex if its algorithmic solution time is growing exponentially in
the network size. At the core of complex {\it optimization} problems
one finds the NP-hard class~\cite{garey79}, where NP stands for
non-deterministic polynomial time.

In the case of uncorrelated networks with power law degree
distributions we can take profit of the existence of hubs to solve
different problems, like destroying the giant
component~\cite{percolation}, preventing epidemic
outbreaks~\cite{spreading}, and searching~\cite{searching}.  The
extremely inhomogeneous structure of uncorrelated networks can also be
exploited to approximate or even to solve instances of NP-hard
problems using heuristic algorithms running in polynomial
time. However, the influence of properties like degree correlations or
clustering is not clear yet. Recent studies of
percolation~\cite{newman02} and disease spreading~\cite{boguna02} have
shown that degree correlations can quantitatively change, e.g., the
transition threshold but qualitatively the results are similar to
those obtained for uncorrelated networks.

This changes drastically if we consider hard optimization tasks defined 
over correlated networks. In this work we study the influence
of degree correlations on the
computational complexity, and in a more general perspective the
relation between the topology of complex networks and the
computational complexity of hard problems defined on top of them. For
this purpose we generalize the Bethe-Peierls approach to
statistical-mechanics models defined on networks with an arbitrary
degree distribution and arbitrary degree correlations of
adjacent nodes. 

The approach is applied to characterize the minimal vertex covers on
these graphs. We have chosen this problem for two reasons: It belongs
to the basic NP-hard optimization problems over graphs \cite{garey79},
and has found applications in monitoring Internet
traffic~\cite{breitbart} and denial of service attack
prevention~\cite{park}. Our analytical results are later compared with
an approximate solution obtained using a heuristic algorithm. This
heuristic fails to find minimal vertex covers in the strongly
correlated case, whereas networks with low correlations seem to be
simple from the point of view of combinatorial optimization.  Within
our analytical approach, this change of behavior is associated with
replica symmetry breaking (RSB).

Consider the set of undirected graphs with $N$ vertices and arbitrary
degree distribution $p_d$. Following a randomly chosen edge, we will
find a vertex of degree $d+1$ with probability $q_d=(d+1)p_{d+1}/c$,
with $c$ denoting the average degree. The number $d$ of additional
edges will be called {\it excess degree}. We further assume 
correlations between adjacent vertices: The probability that a
randomly chosen edge connects two vertices of excess degrees $d, d'$
is given by $(2-\delta_{d,d'})e_{dd'}$. The conditional probability 
that a vertex of excess degree $d$ is reached following any edge 
coming from a vertex of excess degree $d'$,
\begin{equation}
\label{eq:conditioned} 
p(d|d') = e_{dd'}/ q_{d'}\ ,
\end{equation}
thus explicitly depends on both $d$ and $d'$. Consistency with the 
degree distribution requires $\sum_{d'=0}^\infty e_{dd'} = q_d$, 
and $e_{dd'}$ has to be symmetric. For uncorrelated graphs
$e_{dd'}=q_dq_{d'}$ factorizes. The strongest positive correlations 
are reached for $e_{dd'}=q_d\delta_{dd'}$ where only vertices of equal 
degrees are connected.

Let us now consider a general statistical-mechanics model with
discrete degrees of freedom defined on vertices, and interactions
defined on edges.  We use a lattice gas model described by the
Hamiltonian
\begin{equation}
\label{eq:hamiltonian}
-\beta H = \sum_{i<j} J_{ij} w(x_i,x_j) + \mu \sum_i x_i
\end{equation}
defined for any microscopic configuration $x_i=0,1,\ 
i=1,...,N$. $J$ is the adjacency matrix with entries $J_{ij}=1$ if
vertices $i$ and $j$ are adjacent, and $J_{ij}=0$ else. The inverse 
temperature is denoted by $\beta$, the chemical
potential by $\mu$. The interactions $w(x_i,x_j)$ are arbitrary, thus
including also the case of the ferromagnetic Ising model, $w(x_i,x_j)
= (2x_i-1)(2x_j-1)$. The only disorder present in
Eq. (\ref{eq:hamiltonian}) is given by the edges
$J_{ij}$. Generalizations to disordered interactions, as present
e.g. in spin-glasses, random local fields or non-binary discrete
variables are evident.  For clarity of the presentation, we restrict
ourselves to the simple model given above.

Since the graphs are locally tree-like, the model can be solved by the
iterative Bethe-Peierls scheme which becomes exact if only one pure
state is present. The free energy can be expressed in terms of simple
effective-field distributions acting on vertices of given
degree. In the case of multiple pure states this has to be generalized
to the cavity approach, see e.g. \cite{MePa} for the example of a
spin-glass on a Bethe lattice of constant vertex degree.
Alternatively, one can apply the replica approach. The simple
Bethe-Peierls solution corresponds to the assumption of replica
symmetry (RS), whereas the full cavity approach is able to handle also
the case of RSB.

Take any edge $(i,j)$, i.e. $J_{ij}=1$. Let us introduce $Z_x^{(i|j)}$
as the partition function of the subtree rooted in $i$, with deleted
edge $(i,j)$, and with $x_i$ fixed to the value $x$. This partition
function can be calculated iteratively,
\begin{eqnarray}
\label{eq:Ziteration}
Z_0^{(i|j)} &=& \prod_{k\neq j| \ J_{ik}=1 } 
\left( e^{w(0,0)} Z_0^{(k|i)} + e^{w(0,1)} Z_1^{(k|i)} 
\right) \\
Z_1^{(i|j)} &=& e^\mu \prod_{k\neq j| \ J_{ik}=1 } 
\left( e^{w(1,0)} Z_0^{(k|i)} + e^{w(1,1)} Z_1^{(k|i)} 
\right) \nonumber
\end{eqnarray}
The effective fields
$h_{(i|j)} = \ln (  Z_1^{(i|j)} /  Z_0^{(i|j)} )$         
are thus determined by the iterative description
\begin{equation}
\label{eq:fielditeration}
h_{(i|j)} = \mu + \sum_{k\neq j| \ J_{ik}=1 } u( h_{(k|i)} ) \ .     
\end{equation}
where $u( h_{(k|i)} )$ is the effective field induced by $x_k$ on site
$i$, and is given by
\begin{equation}
\label{eq:inducedfield}
u( h_{(k|i)} ) = \ln \left( \frac{ e^{w(1,0)} + e^{ w(1,1)+h_{(k|i)}}
    }{ e^{w(0,0)} + e^{w(0,1)+h_{(k|i)}} } \right) \ .
\end{equation}
The free energy of the system can be written as
\begin{equation}
\label{eq:freeenergy}
- \beta N f = \sum_i f_i + \sum_{i<j} J_{ij} ( f_{ij} - f_i - f_j )
\end{equation}
where the link contribution equals
\begin{equation}
\label{eq:linkf}
 f_{ij}= -\ln\left(
  \sum_{x_i,x_j}e^{w(x_i,x_j)+h_{(i|j)}x_i+h_{(j|i)}x_j} \right)\ ,
\end{equation}
whereas the site contribution
\begin{equation}
\label{eq:sitef}
 f_{i} = -\ln\left( \sum_{x_i} e^{h_i x_i}\right)
\end{equation}
depends on the cavity field
\begin{equation}
\label{eq:cavityfield}
h_i = \mu + \sum_{k | \ J_{ik}=1 } u( h_{(k|i)} )     
\end{equation}
resulting from the influence of {\it all} neighbors on vertex $i$.

Let us now assume, that the model has only one pure state, which
corresponds to the assumption of RS. In this case, the iterative
procedure given by Eq. (\ref{eq:fielditeration}) converges to
well-defined distributions $P_d(h)$ of effective fields $h_{(i|j)}$
restricted to vertices of excess degree $d$. They are determined by
the self-consistency equation
\begin{eqnarray}
\label{eq:pd}
P_d(h) &=& \int_{-\infty}^{\infty} \prod_{l=1}^d \left( dh_l 
\sum_{d_l=0}^\infty p(d_l|d) P_{d_l} (h_l) \right)\nonumber\\
&&\times \delta\left( h - \mu - \sum_{l=1}^d u(h_l) \right)\ .
\end{eqnarray}
Please note that, in contrast to the uncorrelated case, we do need the
field distributions for all possible excess degrees. In the
uncorrelated case, the average of these distributions over $q_d$ is
sufficient. We may also introduce the analogous distributions $\tilde
P_d (h)$ of cavity fields $h_i$ for vertices of given degree $d$
(note: here the full degree is relevant), they can be calculated from
$P_d(h)$ by
\begin{eqnarray}
\label{eq:pdtrue}
\tilde P_d(h) &=& \int_{-\infty}^{\infty} \prod_{l=1}^d \left( dh_l 
\sum_{d_l=0}^\infty p(d_l|d-1) P_{d_l} (h_l) \right)\nonumber\\
&&\times \delta\left( h - \mu - \sum_{l=1}^d u(h_l) \right)\ .
\end{eqnarray}
Replacing the sum over vertices and edges in Eq. (\ref{eq:freeenergy})
by the corresponding averages over $P_d(h)$ resp. $\tilde P_d(h)$,
we finally find the free-energy density
\begin{eqnarray}
\label{eq:f}
\beta f &=& - \sum_{d=0}^\infty (d-1)\ p_d \ \int_{-\infty}^{\infty}
dh \ \tilde P_d(h) \ln \left( 1+e^h \right) \nonumber\\
&&+ \frac c2  \sum_{d,d'=0}^\infty e_{dd'}
\int_{-\infty}^{\infty} dh\ dh' P_{d}(h) P_{d'}(h') \\
&& \ln ( e^{w(0,0)} + e^{w(1,0)+h}+ e^{w(0,1)+h'}
+ e^{w(1,1)+h+h'} )\nonumber
\end{eqnarray}
 
The simplest application of this approach is given by the
ferromagnetic Ising model. If we look to the ground-states, i.e. to
the limit $\beta \to \infty$, we find that, as to be expected, the
global magnetization is determined by the size of the giant
component. The existence of a ferromagnetic phase at low temperature
is thus related to percolation. The latter was already analyzed in
Ref. \cite{newman02}.

Another application is given by the vertex cover (VC) problem. It
belongs to the basic NP-hard optimization problems \cite{garey79}
and, therefore, it is expected to require a solution time which is
growing exponentially with the graph size.  Let us be more
precise. Given a graph with vertices $i\in \{1,...,N\}$ and edges $\{
(i,j) | 1\leq i<j \leq N, J_{ij}=1 \}$, a {\it vertex cover} $V$ is a
subset of vertices, $V\subset \{1,...,N\}$, such that at least one
end-vertex of every edge is contained in $V$. So no edge $(i,j)$ is
allowed to exist with $i\notin V$ and $j \notin V$. Of course, the set
of all vertices forms a trivial VC. The hard optimization
problem consists in finding the {\it minimal VC}.

Using the hard-sphere lattice-gas representation introduced in 
\cite{WeHa}, where $x_i = 1$ if $i\notin V$, and $x_i = 0$ if 
$i\in V$, the VC condition can be rewritten as
\begin{equation}
\prod_{(i,j)| J_{ij}=1} (1-x_i x_j) = 1
\end{equation}
which fits into the above framework by setting $e^{w(x_i, x_j)} =
1-x_i x_j$. The chemical potential can be used to fix the cardinality
of the VC, minimal ones are obtained in the limit $\mu \to
\infty$. They correspond to maximal packings in the lattice-gas
picture. To perform the limit $\mu\to\infty$ all fields have to be
rescaled as $h=\mu z$ \cite{WeHa}. We obtain
\begin{eqnarray}
\label{eq:pdvc}
P_d(z) &=& \int_{-\infty}^{\infty} \prod_{l=1}^d \left( dz_l 
\sum_{d_l=0}^\infty p(d_l|d) P_{d_l} (z_l) \right)\nonumber\\
&&\times \delta\left( z - 1 - \sum_{l=1}^d \max (0,z_l) \right)\ ,
\end{eqnarray}
which is solved by $P_d(z)=\sum_{l=-1}^{+\infty} \rho_l^{(d)}
\delta(z+l)$.  This ansatz allows for integer valued fields only, 
we find a simple relation including only the $\rho_{-1}^{(d)}$:
\begin{equation}
\label{eq:rho1}
\rho_{-1}^{(d)} = \left[ \sum_{d_1=0}^\infty p(d_1|d) 
(1-\rho_{-1}^{(d_1)} ) \right]^d
\end{equation}
All other $\rho_{l}^{(d)}$ follow easily. The expression inside the
parenthesis can be understood as the average probability $\pi_d$ that 
an edge entering a vertex of degree $d+1$ carries a constraint, i.e.
that it is not yet covered by the neighboring vertex. It thus fulfills 
the condition
\begin{equation}
\label{eq:constraint}
\pi_d = \sum_{d_1=0}^\infty p(d_1|d) (1-\pi_{d_1})^{d_1}\ .
\end{equation}
Having in mind that, due to the limit $\mu\to\infty$, every vertex
with positive $z$ is fixed to $x=1$, every one with negative $z$ has
$x=0$, we can immediately read off the fraction of vertices belonging
to a minimal VC,
\begin{equation}
\label{eq:xc}
x_{c} = 1 - \sum_{d=0}^\infty p_d ( 1-\pi_{d-1})^{d-1}
\left( 1 + \frac{d-2}2 \pi_{d-1} \right)\ .
\end{equation}

Remember that the last expressions are related to the validity of RS,
i.e. to the existence of a single connected cluster of minimal VCs in
configuration space. As observed in \cite{WeHa}, RS is related to the
{\it local} stability of this solution. In presence of RSB,
Eq. (\ref{eq:constraint}) has no stable solution. Since it has to be
solved by numerical iteration in the general case, an instability
prevents the program from convergence and thus provides a precise
tool to detect RSB without any RSB calculation.

\begin{figure}
\includegraphics[width=3in]{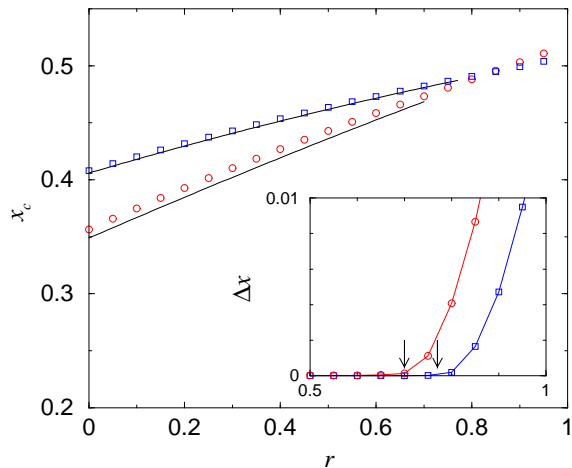}
\caption{
Minimal VC size for a network with degree distribution $p_d\sim
d^{-\gamma}$ and degree correlations given by
Eq. (\ref{eq:assort}). The lines give the analytical solution for
$\gamma=2.5$ (lower curve) and $\gamma=3.0$ (upper curve). The curves
stop at the point where RS breaks.  The symbols are numerical
estimates for $\gamma=2.5$ (circles) and $\gamma=3.0$ (squares), and
network size $N=10^6$. In the inset we plot the upper error bound
$\Delta x$ for generalized leaf removal.  The onset of non-zero error
coincides with the RSB transition, marked by arrows.}
\label{fig:1} 
\end{figure}

To see how this works out, we concentrate on networks having equal
degree distributions but different correlation properties. We restrict
our attention to scale-free graphs with $p_d \sim d^{-\gamma}$ for
$d=1,...,\infty$, with $\gamma > 2$. For vertex cover, interesting
effects are expected to appear for positive correlations, or
assortative networks. We therefore consider
\begin{equation}
\label{eq:assort}
e_{dd'} = q_d \left[ r \delta_{d,d'} + (1-r) q_{d'} \right]\ .
\end{equation}
For $0 \leq r \leq 1$, this expression linearly interpolates between
uncorrelated ($r=0$) and fully assorted ($r=1$) networks. Please note
that this network is percolated for all $\gamma$ as soon as $r>0$. In
Fig. \ref{fig:1} we show the resulting size of the minimal VCs for
different values of $\gamma$ as a function of $r$. The RS solution
breaks at a certain value of $r$ that depends on $\gamma$. There, the
solution-space structure changes drastically, from being unstructured,
or RS, in the low correlated case to being clustered, or RSB, for
sufficiently high correlations.

To check the consequences of this transition for heuristic
optimization algorithms, we have numerically generated scale-free
networks with correlations (\ref{eq:assort}), and applied a
generalization of the leaf-removal heuristic of Bauer and Golinelli
\cite{bauer01}. For the special case of correlations given by
Eq. (\ref{eq:assort}), the network can be generated using a
modification of the Molloy-Reed algorithm \cite{MoRe} for random
graphs of arbitrary degree distribution.  First, each node $i$ is
assigned a degree $d_i$ with probability $p_{d_i}$. Then we create a
set ${\cal L}$ containing $d_i$ copies of each node $i$. Finally,
pairs of nodes are connected according to following rule: {\it (i)}
Select a node $i$ in ${\cal L}$ at random.  {\it (ii)} With
probability $r$, select a second node $j\in {\cal L}$ with $d_j=d_i$
in ${\cal L}$ at random; otherwise select an arbitrary node $j\in
{\cal L}$ at random.  {\it (iii)} Connect $i$ and $j$ and delete both
from ${\cal L}$.  This is repeated until ${\cal L}$ is empty.

Once the network is generated, we construct a VC using a generalized
leaf-removal algorithm \cite{weigt02} defined as follows: Select a
vertex of minimal current degree from the network and cover all its
neighbors. The considered vertices and all incident edges are removed
from the network. This step is iterated untill the full network is
removed. If, for some graph, this algorithm stops without having ever
chosen vertices of current degree $d\geq 2$, the constructed VC is
minimal \cite{bauer01}. Overestimations may appear if the algorithm is
forced to select also vertices of higher degree $d\geq 2$, where the
error can be at most $d-1$. Thus, summing $(d-1)(1-\delta_{d,0})/N$
over all iteration steps, we get an upper bound $\Delta x$ on the
total error made in estimating $x_c$ using the above heuristic
algorithm. If $\Delta x$ goes to zero in the large-$N$ limit,
the algorithm has consequently constructed an almost minimal VC.

In Fig. \ref{fig:1} we show the size of VCs found by generalized leaf
removal as a function of $r$. Up to the RSB point the numerical
solutions are close to the analytical values, up to finite-size
corrections resulting mainly from a degree cutoff $d_{max}\sim
N^{1/\gamma}$. Beyond the RSB point we still have a numerical estimate
but we cannot be sure that it is optimal.  In the inset of
Fig. \ref{fig:1} the upper bound on the error is
displayed. In the RS region we have $\Delta x\approx0$ for $N\gg1$
and, therefore, the heuristic algorithm asymptotically yields the
exact value $x_c$. However, in the highly correlated region we find a
finite $\Delta x$ at any network size, thus the heuristic algorithm
fails to find almost minimal vertex covers. Moreover, the point where
$\Delta x$ becomes different from zero coincides with the RSB point.

It is interesting to know in which phase realistic networks are. VCs have found
applications in monitoring the Internet traffic~\cite{breitbart} and in denial of service
attack prevention~\cite{park}. The analysis of Internet maps has revealed negative
(dissasortative) correlations at the autonomous system level~\cite{pastor01}. Negative
correlations are actually common in technological and biological networks
\cite{newman02}. Hence, the generalized leaf-removal heuristic should output almost
optimal VCs in linear time. On the contrary, social networks exhibit positive
(assortative) correlations~\cite{newman02}. In this case VCs can be used to monitor
social relations between pairs of individuals but, because of the existence of positive
correlations, simple heuristic algorithms may fail to produce near optimal solutions.

To summarize, we have generalized the Bethe-Peierls approach to random
networks with degree correlations and analyzed the VC problem as a
prototype optimization problem defined over graphs. We have found
that uncorrelated power-law networks are simple from the point
of view of combinatorial optimization, inhomogeneities of
neighboring vertices can be exploited. The introduction of
sufficiently large degree correlations leads to RSB and thus
to a failure of simple heuristic algorithms. For constructing optimal 
solutions, complete algorithms including e.g. backtracking have to be 
used. These, however, result in general in exponential solution times, 
and thus in a higher algorithmic complexity. Our results point out that
optimization problems in many technological and biological networks can 
be simple due to the 
strong degree inhomogeneities and negative correlations present on them.

\begin{acknowledgments}
We acknowledge fruitful discussions with M. Leone, A. Vespignani and 
R. Zecchina.
\end{acknowledgments}

\end{document}